\pdfoutput=1

\documentclass{appolb}

\usepackage{amssymb}
\usepackage{hyperref}
\usepackage{graphicx}
\usepackage{amsmath}
\usepackage[numbers,sectionbib,sort&compress]{natbib}

\begin{document}

\title{Partial correlation analysis in ultra-relativistic nuclear collisions%
\thanks{Supported by the Polish National Science Centre grant 2015/19/B/ST2/00937.}
\thanks{Presented by WB at XIII Workshop on Particle Correlations and Femtoscopy (WPCF~2018), Cracow, Poland, 22-26 May 2018.}}

\author{Wojciech Broniowski$^{1,2}$ and Adam Olszewski$^1$
\address{$^1$Institute of Physics, Jan Kochanowski University, 25-406 Kielce, Poland \\
               $^2$The H. Niewodnicza\'nski Institute of Nuclear Physics \\ Polish Academy of Sciences, 31-342 Cracow, Poland}
}
\maketitle

\begin{abstract}
We show that  the method of partial covariance is a very efficient way to introduce constraints 
(such as the centrality selection) in data analysis in ultra-relativistic nuclear collisions. 
The technique eliminates spurious event-by-event fluctuations of physical quantities due to fluctuations of control variables. 
Moreover, in the commonly used superposition approach to particle production the method can be used 
to impose constraints on the initial sources rather than on the finally produced particles, thus separating out the trivial fluctuations 
from statistical hadronization or emission from sources and focusing strictly on the initial-state physics.
As illustration, we  use simulated data from hydrodynamics started on the wounded-quark event-by-event 
initial conditions, followed with statistical hadronization, to show the practicality
of the approach in analyzing the forward-backward multiplicity fluctuations. We mention generalizations to the case with 
several constraints and other observables, such as the transverse momentum or eccentricity correlations. 
\end{abstract}

\bigskip 
  
This talk is based on~\cite{Olszewski:2017vyg} where more details  can be found.  The technique of {\em partial 
covariance} is widely used in other areas of science in situations where one can distinguish the {\em physical}
variables and the {\em control} (spurious, nuisance) variables 
in multivariate statistical samples (see, e.g.,~\cite{Cramer:1946,krzanowski:2000}). 
Such a separation occurs in typical setups in ultra-relativistic nuclear collisions, where response of certain detectors 
is used to determine {\em centrality}, the quantile from a given measure quantity, which plays the role of a control variable, whereas other quantities 
correspond to physical variables. 

The problem of eliminating fluctuations of centrality, spuriously correlating to physical quantities, has a long history with numerous
methods, e.g.~\cite{Gazdzicki:1992ri,Gorenstein:2011vq,Bhalerao:2014mua,Broniowski:2017tjq,Rogly:2018kus}, developed 
precisely for this purpose. We argue that the advocated partial covariance method is particularly simple, bringing up
a general understanding of centrality as a control variable, whose interpretation depends on the experimental 
arrangement.

\begin{figure}[b]
\begin{center}
\includegraphics[width=0.44\textwidth]{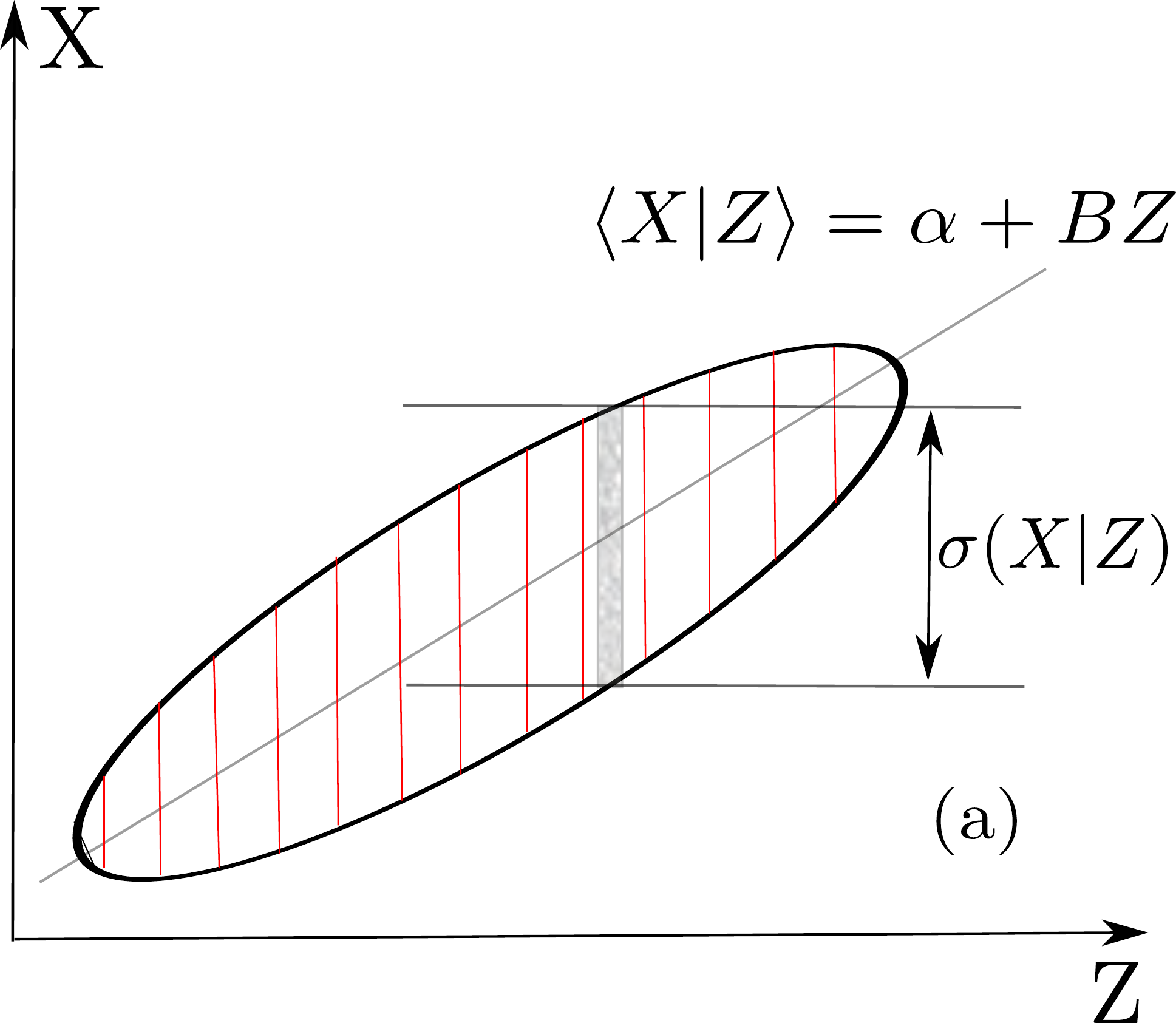}  \hfill \includegraphics[width=0.44\textwidth]{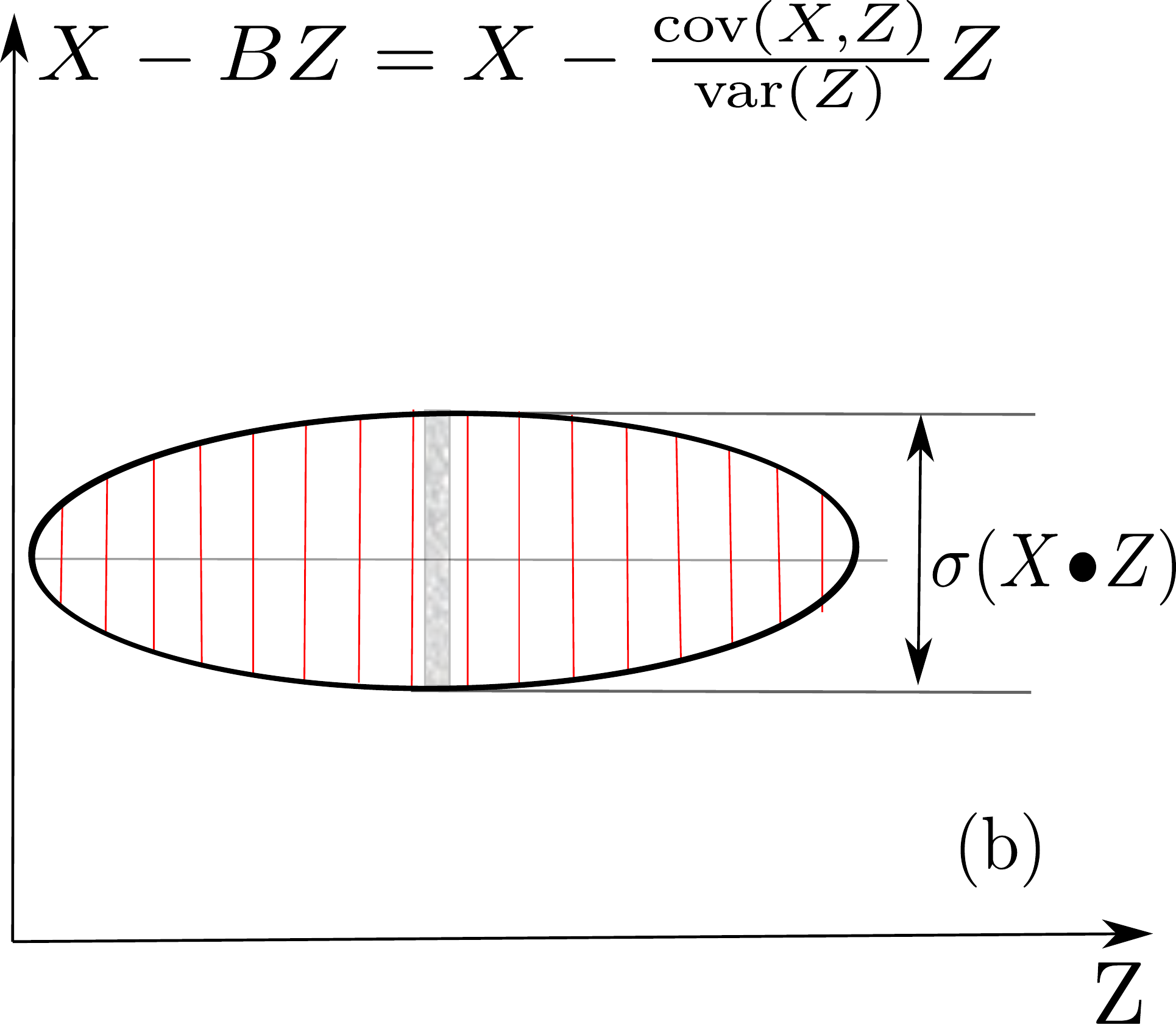} 
\caption{Graphical illustration of the proof of equality of the partial and conditional variance.    \label{fig:cond}}
\end{center}
\end{figure}

With partial correlations, the constraints on the control variables emerge from the 
relationship of the partial covariance, defined as
\begin{eqnarray}
{\rm cov}(X_i,X_j \bullet Z) \equiv  {\rm cov}(X_i,X_j)- {\rm cov}(X_i,Z_a) \left [ {\rm cov}(Z)\right ]^{-1}_{ab} {\rm cov}(Z_b,X_j), \label{eq:pcov}
\end{eqnarray}
to the {\em conditional covariance} ${\rm cov}(X_i,X_j | Z)$, where $i,j$ label the physical variables $X$  and $a,b$ the control variables $Z$. 
The subtraction in Eq.~(\ref{eq:pcov}) projects out the components spuriously correlated to the control variables, one is thus left with 
physical correlations only.
The conditional covariance is 
defined by first fixing the values of $Z$ to obtain the covariance of $X_i$ and $X_j$, and then averaging over the control variable(s) $Z$.
Note that the prescription of using very narrow centrality bins and then averaging over them, 
advocated in~\cite{Abelev:2009ag,Bzdak:2011nb,De:2013bta}, precisely conforms to this recipe. 
It has been shown~\cite{Lawrance:1976,Baba:2004,Bzdak:2011nb} that ${\rm cov}(X_i,X_j \bullet Z)={\rm cov}(X_i,X_j | Z)$ iff 
$\langle X | Z\rangle = \alpha + B Z$, i.e., iff the expectation value of $X$ at $Z$ fixed is an affine function of $Z$, with $\alpha$ 
a constant vector and $B$ a constant matrix. This feature is well satisfied when the centrality classes are sufficiently narrow, 
as is typically the case, hence the 
equality between conditional and partial correlations holds to a very good approximation.

To understand in simple terms this relation, 
let us consider the case with a single physical variable $X$ 
and a single constraint $Z$. In panel (a) of Fig.~\ref{fig:cond} we show a data sample (represented with an oval) which is cut into narrow stripes 
with fixed  values of $Z$. By assumption, the expectation value of $X$ in each stripe  aligns along a straight line. The typical width 
at a fixed $Y$ is indicated with $\sigma(X | Y)$. Note that it is much narrower than the whole span of the sample in $X$, which 
is also due to the extension of the sample in $Y$, which is correlated with $X$. 
Now, averaging the affine relation over $Z$ yields $B= {\rm cov}(X,Z)/{\rm var}(Z)$. We may thus 
lower each stripe from panel (a) of Fig.~\ref{fig:cond} by $B Y$, with the result shown in panel (b), where the data oval is ``straightened. 
The variance of this representation of the sample is
${\rm var}(X-BY)={\rm var}(X)-{\rm cov}(X,Y)^2/{\rm var}(Y)$, which is nothing but the partial variance of $X$, which completes the proof
in this simple case. 
The derivation from right to left 
proceeds analogously. 

The construction of Fig.~\ref{fig:cond} indicates the two equivalent ways to evaluate the covariance of the physical variables: slicing into narrow bins and 
computing the conditional covariance, or computing the partial covariance on the whole sample. We note that the partial covariance method is simpler, as 
it avoids possible problems encountered for low multiplicity samples, where narrow bins may be poorly populated, or even empty. 

We also wish to remark here that the meaning of centrality in nuclear collisions is by no means universal, but 
is strictly related to the response of the chosen ``control'' detectors, be it multiplicity in the central or peripheral bins, 
multiplicity of a sub-event sample, multiplicity of spectators in peripheral detectors, 
or the transverse energy from calorimeters. Definition (\ref{eq:pcov}) allows one
to combine several control variables simultaneously in a straightforward way.

A novel development presented in~\cite{Olszewski:2017vyg} was the combination of the technique of partial covariance with 
the superposition approach~\cite{Olszewski:2013qwa} to particle production in ultra-relativistic nuclear collisions, based on the 
assumption that particles are emitted from independent sources, such as wounded 
nucleons~\cite{Bialas:1976ed} or quarks~\cite{Bialas:1977en,Anisovich:1977av}. The partial covariance method 
makes it possible to impose constraints {\em at the level of these initial sources}, which we find nontrivial. We have tested our formalism by analyzing the 
forward-backward (FB) multiplicity correlations, defined via the correlation function
\begin{eqnarray}
C(X_1,X_2)=\frac{{\rm cov}(X_1,X_2)}{\langle X_1 \rangle \langle X_2 \rangle}.
\end{eqnarray}
The FB correlation of the numbers of sources, with the constraint imposed at the number of sources in 
the mid-rapidity bin $C$, is given by the formula~\cite{Olszewski:2017vyg}
\begin{eqnarray}
C(S_F,S_B\bullet S_C) &=&  {C}(N_F,N_B)
-\frac{{C}(N_F,N_C) {C}(N_B,N_C)}{{\rm var}(N_C) - \omega(m) \langle N_C \rangle} \nonumber \\
&\simeq& C(S_F,S_B | S_C),  \label{eq:p}
\end{eqnarray}
where $\omega(m)$ is the scaled variance of the overlaid distribution ($m$ is a random number of particles emitted from a single source). 
In the case of the Poisson distribution, $\omega(m)=1$. Note that except for $\omega(m)$, 
the quantities on the right-hand side of Eq.~(\ref{eq:p}) involve measurable 
particle multiplicities only.

\begin{figure}[tb]
\begin{center}
\includegraphics[width=0.49\textwidth]{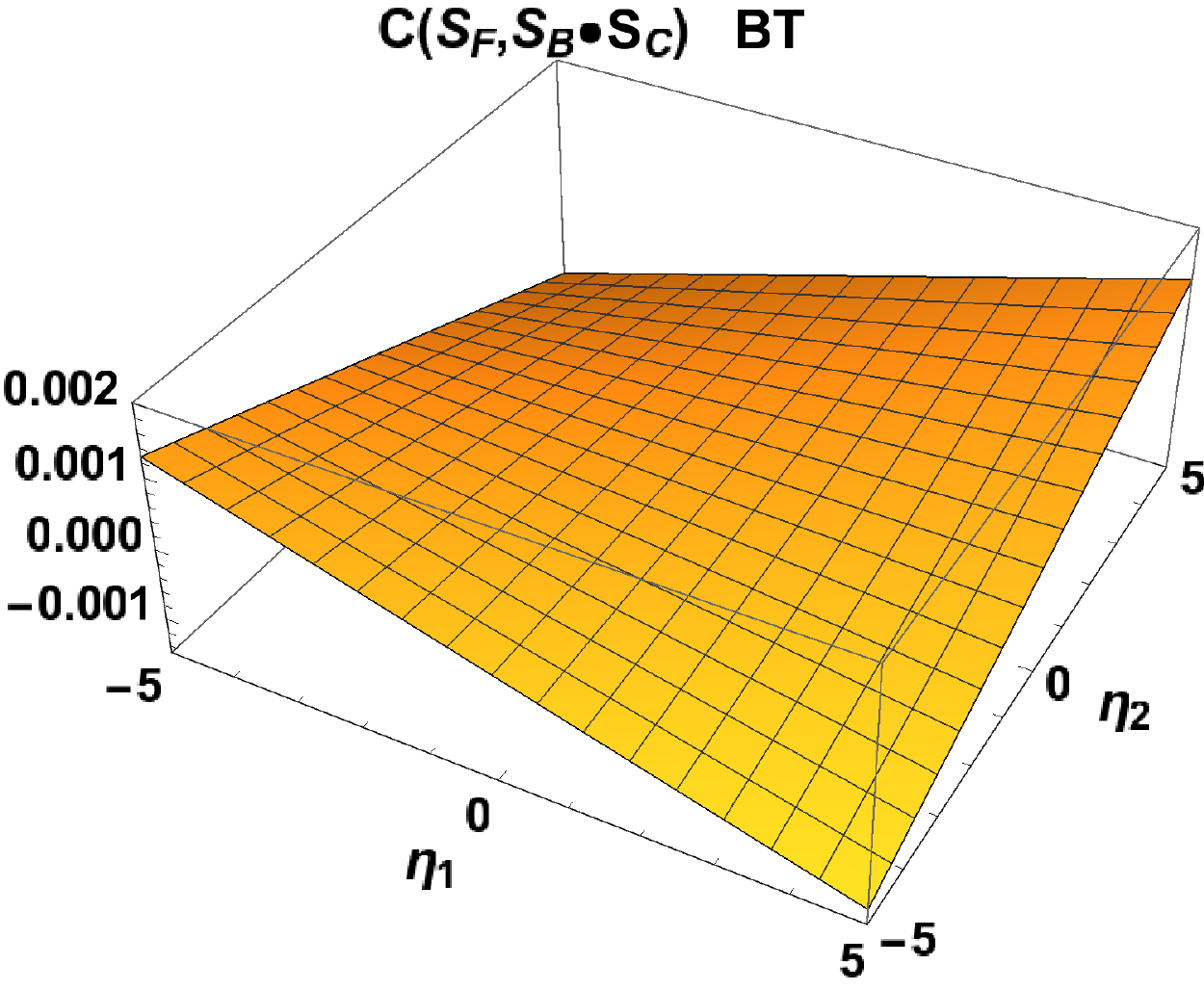} \hfill \includegraphics[width=0.49\textwidth]{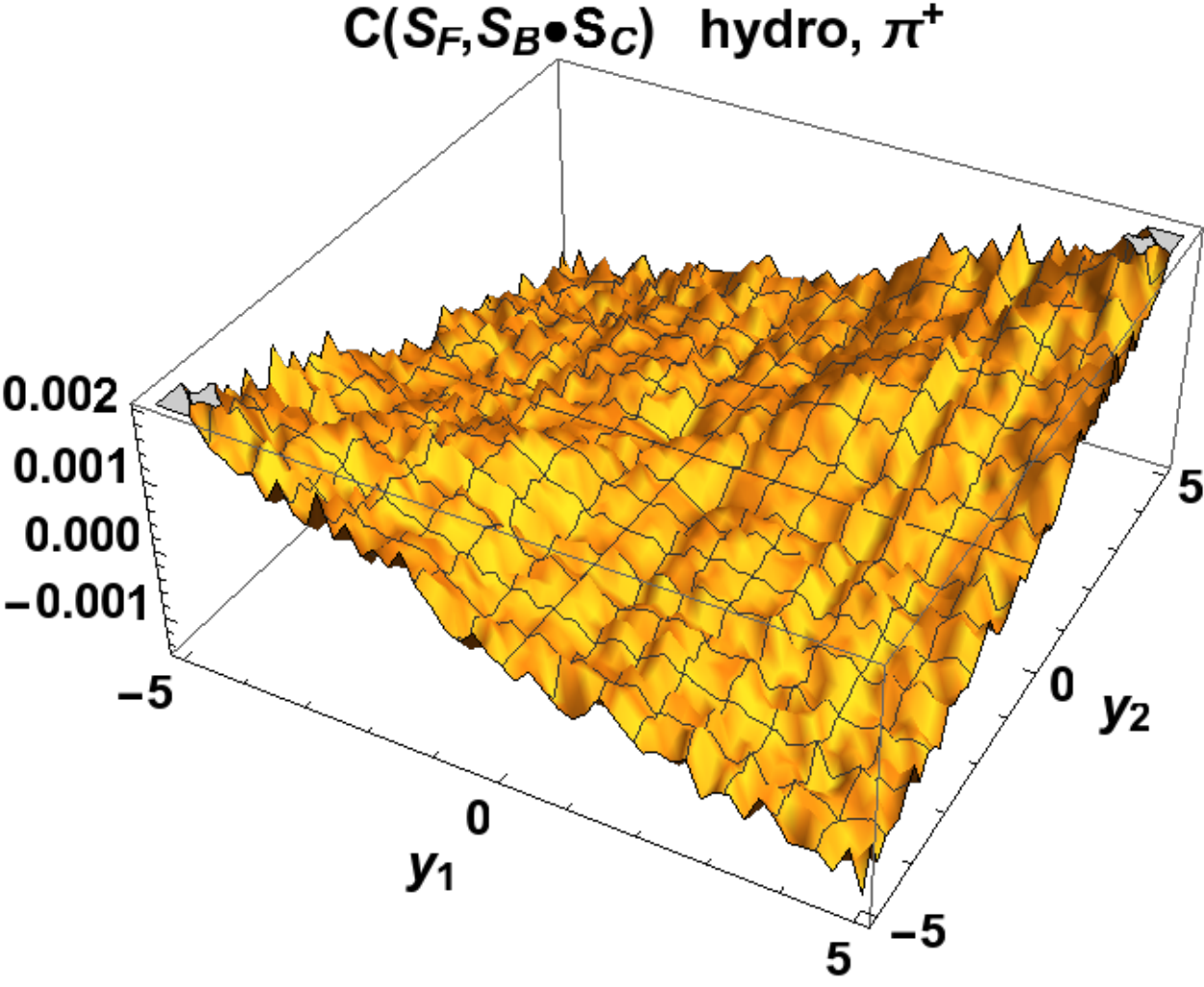}
\caption{Forward-backward partial correlations for the source multiplicities with the constraint fixing the number of sources in the mid-rapidity bin, 
evaluated directly in the BT model~\cite{Bzdak:2012tp} (a), and  
extracted from the simulated data for positively charged pions (b), plotted against the rapidities of the pions, $y_{1,2}$.
\label{fig:pcov}}
\end{center}
\end{figure}

To test formula~(\ref{eq:p}) in a practical application, we have taken a sample of simulated events which uses 
the wounded quark event-by-event initial conditions~\cite{Bozek:2016kpf}, the 3+1D viscous 
hydrodynamics~\cite{Bozek:2009dw}, and {\tt THERMINATOR} for statistical hadronization~\cite{Kisiel:2005hn,Chojnacki:2011hb}.
The longitudinal profile in spatial rapidity is assumed to have the phenomenologically 
successful form of ``triangles''~\cite{Bialas:2004su,Adil:2005bb,Bozek:2010bi}, where each source has the 
emission profile peaked in the direction of its motion.
For our test, on the one hand we have computed the right-hand side with the generated hadrons, on the other hand we have evaluated 
the left-hand side directly from the known correlation of sources. In the adopted Bzdak-Teaney  (BT) model~\cite{Bzdak:2012tp} it has the form
\begin{eqnarray}
C(S_F,S_B\bullet S_C)= \frac{{\rm var}(Q_A-Q_B)}{\langle{Q_A+Q_B}\rangle^2} \frac{y_1 y_2}{y_b^2}, \label{eq:bt}
\end{eqnarray}
where $Q_{A,B}$ denotes the number of wounded quarks in the colliding nucleus $A$ or $B$ and $y_b$ stands for the rapidity of the beam.

As demonstrated in Fig.~\ref{fig:pcov}, we are able to recover the FB partial 
correlations in initial condition to a very reasonable accuracy. The left panel shows the 
partial covariance from the BT model, whereas the right panel presents its estimate extracted with multiplicities of positively charged 
pions from the simulated events (the use of same charge pions reduces the correlations from resonance decays). 
We note a remarkably similar shape
and magnitude of the two results. Some discrepancy may be attributed to breaking of the 
independence of sources, as well as from the mixing of neighboring rapidity bins, 
here coming from cascades of resonance decays. 

Needless to say, it would be very interesting to 
apply the proposed method to real data and to infer information on correlation between multiplicities of 
the initial sources. Generalizations to multiple simultaneous 
constraints (such as from response of various detectors controlling centrality) has been discussed in~\cite{Olszewski:2017vyg}.
The method can also be straightforwardly extended to FB correlations of 
other observables, such as the transverse momentum~\cite{th} or the harmonic flow coefficients.

We thank Piotr Bo\.zek for providing a sample from hydrodynamic simulations in the wounded quark model, 

\bibliographystyle{apsrev4-1}
\bibliography{hydr}

\end{document}